\documentclass[twocolumn]{aastex631}

\usepackage{aas_macros}

\usepackage{amsmath}

\usepackage{my_astro}

\usepackage[all]{hypcap} % makes figure ref take you to figure and not caption

\newcommand{\fuv}{$FUV_{130}$}
\newcommand{\sys}{GJ~436}
\newcommand{\planet}{GJ~436~b}
\newcommand{\alf}{Alfv\'{e}n}
\renewcommand{\deg}{$^\circ$}
\newcommand{\mdeg}{^\circ}
\newcommand{\Caiihk}{\ion{Ca}{2}~H~\&~K}

\begin{document}

% key things for title: SPI hypothesis, flares, cycles, rotation, FUV, GJ 436
\title{Flares, Rotation, Activity Cycles and a Magnetic Star-Planet Interaction Hypothesis for the Far Ultraviolet Emission of GJ 436}

\correspondingauthor[0000-0001-5646-6668]{R. O. Parke Loyd}
\email{astroparke@gmail.com}

\author[0000-0001-5646-6668]{R. O. Parke Loyd}
\affiliation{Eureka Scientific, 2452 Delmer Street Suite 100, Oakland, CA, 94602-3017}

\author[0000-0002-5094-2245]{P. C. Schneider}
\affiliation{Hamburger Sternwarte, 
Gojenbergsweg 112, 21029,
Hamburg, Germany}
\affiliation{Scientific Support Office, Directorate of Science, European Space Research and Technology Center (ESA/ESTEC), Keplerlaan 1, 2201, AZNoordwijk, The Netherlands}

\author[0000-0003-0711-7992]{James A. G. Jackman}
\affiliation{School of Earth and Space Exploration, Arizona State University, 781 E. Terrace Mall, Tempe, AZ 85287, USA}

\author[0000-0002-1002-3674]{Kevin France}
\affiliation{Laboratory for Atmospheric and Space Physics, 
Boulder, CO 80309}

\author[0000-0002-7260-5821]{Evgenya L. Shkolnik}
\affiliation{School of Earth and Space Exploration, Arizona State University, 781 E. Terrace Mall, Tempe, AZ 85287, USA}

\author[0000-0003-2631-5265]{Nicole Arulanantham}
\affiliation{Space Telescope Science Institute, 3700 San Martin Drive, Baltimore, MD 21218, USA}

\author[0000-0001-9207-0564]{P. Wilson Cauley}
\affiliation{Laboratory for Atmospheric and Space Physics,
Boulder, CO 80309, USA}

\author[0000-0003-4450-0368]{Joe Llama}
\affiliation{Lowell Observatory,
1400 West Mars Hill Road,
Flagstaff, AZ 86001, USA}

\author[0000-0002-6294-5937]{Adam C. Schneider}
\affiliation{United States Naval Observatory, Flagstaff Station, 10391 West Naval Observatory Road, Flagstaff, AZ 86005, USA}
\affiliation{Department of Physics and Astronomy, George Mason University, MS3F3, 440 University Drive, Fairfax, VA 22030, USA}
 
\begin{abstract}
Variability in the far ultraviolet (FUV) emission produced by stellar activity affects photochemistry and heating in orbiting planetary atmospheres.
We present a comprehensive analysis of the FUV variability of GJ 436, a field-age, M2.5V star ($P_\mathrm{rot}\approx44~d)$ orbited by a warm, Neptune-size planet  ($M \approx 25$~\Mearth, $R \approx 4.1$~\Rearth, $P_\mathrm{orb}\approx2.6$~d).
Observations at three epochs from 2012 to 2018 span nearly a full activity cycle, sample two rotations of the star and two orbital periods of the planet, and reveal a multitude of brief flares.
Over 2012-2018, the star's $7.75\pm0.10$~yr activity cycle produced the largest observed variations, $38\pm3$\% in the summed flux of major FUV emission lines.
In 2018, variability due to rotation was $8\pm2$\%.
An additional $11\pm1$\% scatter at 10~min cadence, treated as white noise in fits, likely has both instrumental and astrophysical origins.
Flares increased time-averaged emission by 15\% over the 0.88~d of cumulative exposure, peaking as high as 25$\times$ quiescence. 
We interpret these flare values as lower limits given that flares too weak or too infrequent to have been observed likely exist. 
\sys's flare frequency distribution (FFD) at FUV wavelengths is unusual compared to other field-age M~dwarfs, exhibiting a statistically-significant dearth of high energy ($>4\sn{28}$~erg) events that we hypothesize to be the result of a magnetic star-planet interaction (SPI) triggering premature flares. 
If an SPI is present, \planet's magnetic field strength must be $\lesssim$100~G to explain the statistically insignificant increase in orbit-phased FUV emission.
\\
\\
\textbf{Erratum pending:} Due to an arithmetic error, the published limit on the magnetic field strength is incorrect. The correct limit is $\lesssim$10~G.

\end{abstract}


\section{Introduction} 
\label{sec:intro}

GJ~436 is a nearby M2.5V, 0.45~\Msun\ star \citep{hawley03,knutson11} well known for its lone Neptune-size planet ($M = 25.4^{+2.1}_{-2.0}$~\Mearth, $R = 4.10\pm0.16$~\Rearth; \citealt{butler04,gillon07,lanotte14}) that is actively losing atmospheric mass \citep{kulow14,ehrenreich15}.
This planet is an upcoming target for atmospheric characterization by the James Webb Space Telescope (GTO programs 1177 and 1185, PI Greene), and the system has undergone extensive spectroscopic observation at far ultraviolet (FUV) wavelengths \citep{france13,france16,santos19}.
These FUV observations provide an opportunity to investigate the variability of a field-age \citep{torres08}, partially-convective M~dwarf in high-energy emission across a wide range of timescales, from seconds to years.
The data can probe variability as short lived as stellar flares and as long lived as stellar activity cycles. 
This information is relevant to exoplanets, where variations in FUV irradiation affect rates of photochemistry and heating  in their atmospheres \citep[e.g.,][]{segura07}.
Because M~dwarfs are numerous and favorable for exoplanet observations \citep[e.g.,][]{shields16}, the constraints on FUV variability available for \sys\ are broadly applicable to a large number of planets. 

The \sys\ system could be experiencing an interaction between the star and its single close-in planet.
Many forms of star-planet interactions have been proposed, such as tidal suppression of activity, tidal spin up, and shrouding by escaped planetary gas \citep{cuntz00,fossati13, pillitteri14, poppenhaeger14}. 
Of particular interest is a direct magnetic interaction between the planet and host star that results in energy dissipation at the stellar surface, producing a hot spot that circumnavigates the star at the orbital period of the planet \citep{saur13}.
For GJ~436, this is 2.64~d \citep{bourrier18}.
Direct magnetic interactions themselves have many flavors, including particle precipitation from reconnection of stellar and planetary fields, \alf\ wave dissipation, and triggering reconnection of stellar fields \citep{lanza15}.
For magnetic disturbances to reach the star, the \alf\ Mach number at the planet must be $<1$ \citep{saur13}.
Though difficult to predict, this could well be the case for GJ~436~b (see Section \ref{sec:spi}).

A number of past works have identified evidence of this kind of star-planet interaction (SPI) in hot Jupiter systems \citep[e.g.,][]{shkolnik08,walker08,pagano09,cauley19}.
Because the strength of the magnetic SPI, and hence the amplitude of its periodic signal, is controlled by the planetary magnetic field strength, this possibility provides the prospect of probing the magnetic field of \planet\ \citep{lanza15}.
With this in mind, we conducted targeted observations in 2017 and 2018 to sample the star's FUV emission across the planetary orbital period.

In this paper, we report on a comprehensive variability analysis of \sys's FUV emission and a search for a magnetic SPI. 
This work provides an independent replication of a similar analysis conducted as part of the transit study of \cite{santos19} (hereafter dS19).
Their transit analysis confirmed the stability of the planet's remarkable \lya\ transit and demonstrated no significant transit in metal lines.
Meanwhile, their variability analysis yielded evidence for rotational modulation in \Cii, \Siiii, and \Nv\ emission stable over several years as well as magnetic cycle variations leading to optical, \Caii, and FUV variations consistent with changing spot coverage.
In comparison to dS19, the present work adds a population study of the star's flares, details of variability fits, and the results of an SPI search.

\section{Methods}

We analyzed all available observations of GJ~436 made with the Cosmic Origins Spectrograph (COS) aboard the Hubble Space Telescope (HST) that utilized the G130M grating.
The observations originate from four programs that observed during three separate epochs spanning 5.5 years:  HST-GO-12464 (PI France), 13650 (France), 14767 (Sing), and 15174 (Loyd).
The data are aggregated under DOI \dataset[10.17909/6p65-wg08]{https://mast.stsci.edu/portal/Mashup/Clients/Mast/Portal.html?searchQuery=\%7B\%22service\%22:\%22DOIOBS\%22,\%22inputText\%22:\%2210.17909/6p65-wg08\%22\%7D}.
Details of the observations are available at the DOI link where they can be downloaded.

\begin{figure*}
    \includegraphics{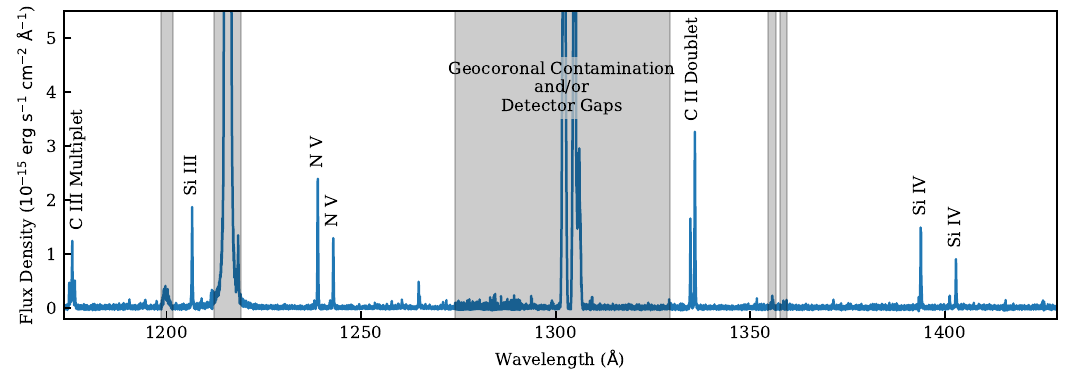}
    \caption{Coadded spectrum of GJ~436. 
    The range of the plot matches that of the $FUV_{130}$ band.
    The gray regions delineate regions of airglow contamination and a detector gap that falls near 1300~\AA\ that we masked from the $FUV_{130}$ band.
    Labels indicate strong emission lines that we also analyzed for variability, and the regions between the labeled lines and gray bands represent the pseudocontinuum. 
    }
    \label{fig:spectrum}
\end{figure*}

\begin{figure*}
    \includegraphics{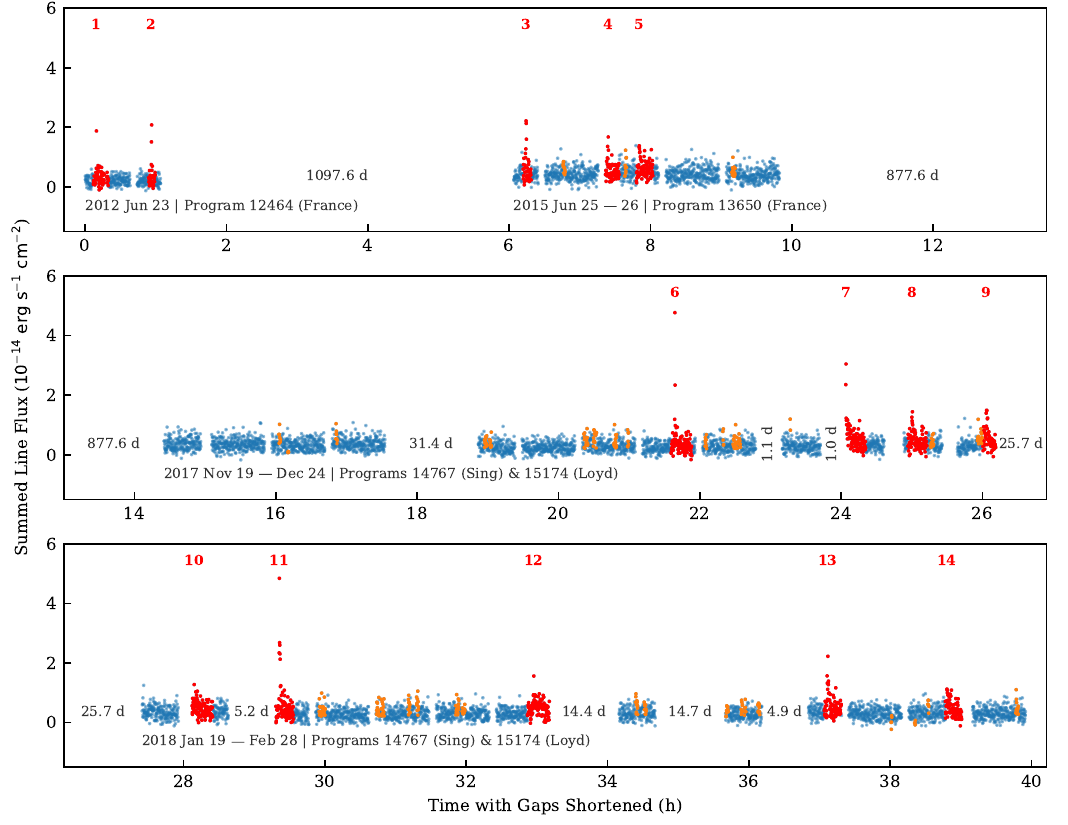}
    \caption{Lightcurve of the summed emission line flux at a 10~s cadence.
    We shortened gaps in the data to a logarithmic scaling of their original lengths as labeled on the plot.
    Labels below each epoch of data indicate the date range covered and the program(s) from which the observations originate.
    Red points indicate flares detected at $>5\sigma$, with assigned labels above them in red.
    Orange points indicate anomalous ($>3\sigma$) runs of data excluded when estimating the quiescence for the purposes of flare detection. 
    }
    \label{fig:overview}
\end{figure*}

The G130M grating covers wavelengths $1150-1450$~\AA\ at a resolving power of 12,000-16,000.
The detector is a photon counter, yielding event lists that can be binned arbitrarily in wavelength and time within the limits of the detector resolution.
From these lists, we created flux-calibrated lightcurves following the methodology presented in \cite{loyd14} and \citeauthor{loyd18b} (\citeyear{loyd18b}; hereafter L18). 
Figure~\ref{fig:spectrum} shows the observed band and Figure~\ref{fig:overview} indicates the quantity and duration of the data originating from the various observing programs. 

We isolated flux within several bands to analyze for variability.
The largest of these is the $FUV_{130}$ band used in the flare analysis of M~dwarf stars by L18.
Figure~\ref{fig:spectrum} plots the spectrum of GJ~436 within the $FUV_{130}$ band, created by coadding all exposures.
As in L18, regions masked in gray were excluded from the band because they are contaminated by geocoronal airglow emission and, in the case of the mask near 1300~\AA, are also affected by a gap between the two butted detectors used by COS. 
The coadded spectrum shown in Figure~\ref{fig:spectrum} does not exhibit this gap because it was eliminated by dithering.
We also analyzed emission from $\pm100$~km~\pers\ bands covering five strong emission lines and multiplets, \Ciii, \Siiii, \Nv, \Cii, and \Siiv, labeled in Figure~\ref{fig:spectrum}.
These lines cover a formation temperature range of 4.5-5.2 in $\log_{10}(\mathrm{K})$, within the stellar transition region \citep{dere97,zanna21}.
For multiplets, we summed flux from all components.
As a final band, we summed flux between the lines, labeling this ``psuedocontinuum'' since it is comprised of both continuum sources and weak or undetected emission lines. 

We identified flares in the data using the method described in L18, with some modifications.
The algorithm identifies flares by isolating runs (series) of points above the mean value of the lightcurve that have an integrated energy at least 5$\sigma$ larger than the average run.
It uses a maximum likelihood estimate for the mean, masking any previously identified flares. 
Chunks of data with gaps longer than 24~h are processed separately, allowing for, e.g., variations in the mean due to rotation. 
The process of estimating the mean and identifying flares is iterated until convergence.
Newly identified flares are masked with each iteration to mitigate the upward bias in the mean produced by flares.

Rather than using the \fuv\ flux to identify flares as in L18, we used the summed emission line fluxes.
These lines are more sensitive to flares than the pseudocontinuum portions of the \fuv\ band.
Including the pseudocontinuum added substantial noise without substantially increasing the signal, reducing the sensitivity to flares.
We also increased the span of data masked around a flare to start 120~s ahead of and extending to 3 times the length of the data identified as a flare. 

After identifying flares in the summed line flux, we measured the equivalent duration, energy, and peak flux of each flare in the \fuv\ band to enable direct comparisons to the results of L18. 
Equivalent duration is the time the star would have to spend in quiescence to emit the same energy in the same band as was emitted by the flare.

Having cleaned the data of flares, we binned the remaining data to $\approx 10$~min and simultaneously fit for variability due to stellar activity cycles, rotation, a possible SPI , and a jitter term.
We only used the 2017/2018 epoch of data to constrain rotation and SPI signals because these were the only data that sampled across the planetary orbital period and stellar rotation period without large gaps where phase and amplitude could have shifted.
We did not mask or otherwise account for planetary transits captured by the 2015 and 2017/2018 epochs because there is no evidence of a transit in the FUV emission lines that we analyzed (\citealt{loyd17,lavie17}, dS19).
We used the MCMC code \texttt{emcee}\footnote{\url{https://emcee.readthedocs.io/en/stable/}} \citep{foreman13} to sample the posterior distribution of the fit parameters using 100 walkers to a factor of at least 100$\times$ the autocorrelation length for the samples of each parameter, taken as the median of the values for all walkers.

The stellar rotation and activity cycle models were simple sine functions allowed to vary in amplitude and phase but fixed to a rotation period of 44.09~d and a cycle period of 7.75~yr as measured from the optical data (described below).
We fit this model to the data integrated to 10~min bins with flares masked. 
After fitting the summed line fluxes, we used the posterior on phase from that fit as a prior for the other, lower SNR bands to better constrain their variability amplitude.

The jitter term was Gaussian white noise added in quadrature to the measurement uncertainty of each data point.
This quantity effectively represents how much more scatter was present in the lightcurve than was expected from measurement uncertainties and is analogous to the excess noise metric of \cite{loyd14}.
However, we used longer time bins of 10~min instead of the 1~min bins used by \cite{loyd14} to offset the faintness of GJ~436, which was the fifth faintest star among the 42 analyzed in \cite{loyd14}.
We allowed the standard deviation of the added noise to vary between epochs, resulting in three separate white noise parameters. 

To fit for a possible magnetic star-planet interaction signal, we assumed a truncated sine model as done for \Caii~H~\&~K in \cite{shkolnik03,shkolnik08}.
This model approximates the appearance of a bright, optically-thick spot of constant flux with a viewing angle that shifts as it traverses the visible stellar hemisphere.
However, FUV line emission within a hot spot might not be optically thick.
If emission originates from an optically thin slab of plasma, and is emitted isotropically, then the flux would be constant once the slab emerged from behind the stellar disk (e.g., \citealt{toriumi20}).
To allow for this possibility, we added a shape parameter, $\alpha$, to the truncated $\sin$ model, yielding a model of the form
\begin{equation}
F = 
\left\{\begin{matrix}
0, & \sin\phi\leq0\\ 
\sin^\alpha\phi, & \sin\phi > 0
\end{matrix}\right.,
\end{equation}
where 
\begin{equation}
    \phi = \frac{2\pi}{P_\mathrm{orb}}t + \Phi,
\end{equation}
$A$ is amplitude, $P_\mathrm{orb}$ is the planetary orbital period of 2.64~d \citep{bourrier18}, and $\Phi$ is a phase offset.
As $\alpha\rightarrow0$, the function approaches a top hat, representative of the case where the hot spot emission is optically thin and the observed flux simply depends on whether the spot is visible or not.
This ad hoc parameterization allowed the code to explore sinusoidal, top hat, and intermediate solutions.
We fit this model to the data with flares included because they could be triggered by an SPI, simultaneous with rotation and cycle fits but with a separate jitter term. 
An initial run indicated only upper limits would result.
These limits could be unreasonably large if we allowed phase to vary freely since the sampler could place the signal peak in a data gap.
To provide a reasonable upper limit, we fixed the signal to be in phase with the orbit. 
Although phase offsets likely occur for an SPI-induced hot spot, they cannot be easily predicted \citep{lanza13}, and we consider a fixed-phase fit reasonable for an order-of-magnitude upper limit. 

\begin{figure*}
    \includegraphics{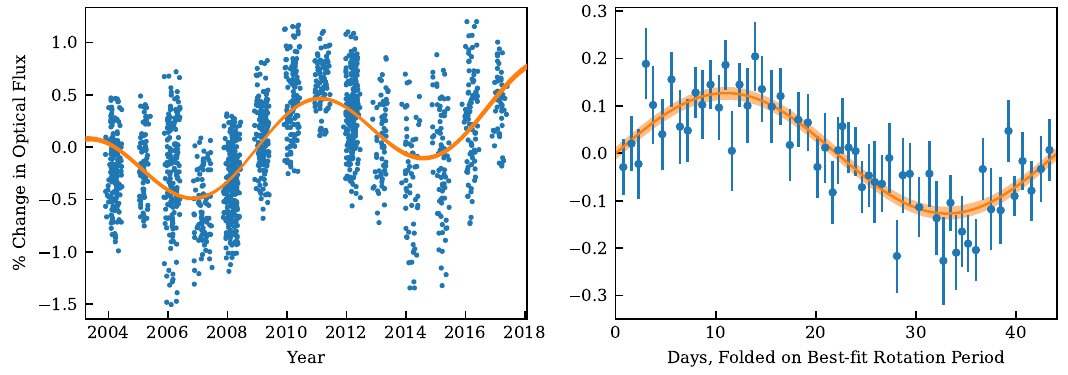}
    \caption{
    Our fit to the Str\"omgren $(b+y)/2$ photometry from \cite{lothringer18}, mimicking their Figure~1.
    Left: Data with the sinusoidal fit to the $7.75\pm0.10$~yr activity cycle period and the linear trend overlaid.
    Line thickness represents uncertainty in the fit.
    Right: Data with the activity cycle sinusoid and linear trend subtracted, then folded onto the best fit rotation period of $44.09\pm1.16$~d, and binned by groups of 30.
    The rotation model sinusoid is overlaid with the translucent region representing uncertainty in the fit. 
    Error bars are the sample error on the mean of the binned points. 
    }
    \label{fig:lothringer}
\end{figure*}

To provide a consistent comparison to the optical variability of GJ~436, we also analyzed the photometry in the Str\"omgren $b$ and $y$ band filters described in \cite{lothringer18}.
We used the same techniques outlined above, but included rotation period and the slope and intercept for a linear rise in flux as additional free parameters.
In contrast with \cite{lothringer18}, we fit the $(b + y)/2$ photometry in flux rather than magnitude space.
We set priors for the period of the rotation sinusoid to be $< 365$~d, the period of the activity cycle sinusoid to be $> 365$~d, and all amplitudes to be $\geq 0$.
These priors did not prove restrictive. 
Figure~\ref{fig:lothringer} shows the results of this fit.  

\section{Results \& Discussion}
\subsection{Flares}
\label{sec:flares}

We identified 14 flares in the data, shown in the context of the entire dataset in Figure~\ref{fig:overview}.
Table~\ref{tbl:flares} gives the properties of these flares in the \fuv\ band, enabling a comparison to L18, as well as the summed line flux used to identify the flares and shown in the figures.
Most flares exhibit a distinct peak.
Event 7 (and possibly event 10) appears to be the decay phase of a flare that began prior to the start of an exposure.
Events 5 and 14 do not exhibit clear peaks and, despite passing the statistical cut, could be false positives.
We cannot definitively explain these flux increases.
One possibility is that they are a manifestation of magneto-acoustic waves or wave interference such as produce minute-timescale variations in transition-region emission on the Sun, though it is unclear if these variations significantly modulate disk-integrated emission \citep{sangal22}.
Similar variations with a greater amplitude that stands out well above the noise were observed on a young M star by \cite{loyd18a}.
Another possible explanation is that these anomalies are conglomerations of weak flares. 
 % note that in the smoothed flux there are several quiescence crossings prior to the peak of flare 10, but overall trend does look like a decay

\begin{figure*}
    \includegraphics{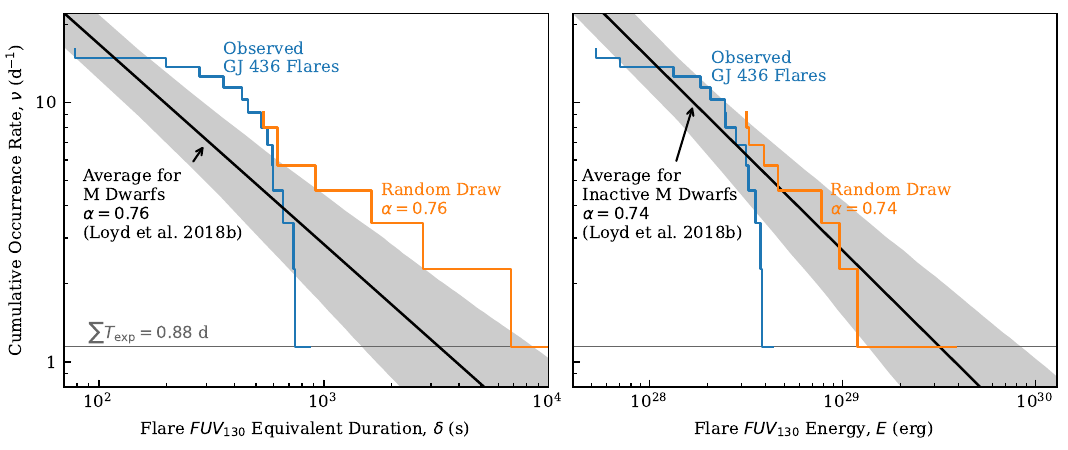}
    \caption{
    Cumulative frequency distribution of GJ~436 flares in the \fuv\ band.
    Gray regions give the uncertainty envelope from fits to a compendium of M~dwarf flares in L18.
    To highlight the cliff in the GJ~436 flares, the figure also shows random draws of even equivalent durations and energies assuming the same number of events above 500~s and $2.5\sn{28}$~erg and a inverse power-law slope, $\alpha$, equal to the median observed by L18.
    The gray horizontal line represents the limit where one event would be expected to occur within the cumulative observing time.}
    \label{fig:ffd}
\end{figure*}

The flares exhibited by GJ~436 are frequent, and yet they are of unusually low energy.
Figure~\ref{fig:ffd} shows the cumulative flare frequency distribution (FFD) of the observed flare energies and equivalent durations (a metric of flare energy that normalizes by the star's quiescent emission) in comparison with FFDs made by combining observations of 10 M~dwarfs in the same \fuv\ band (L18).
Although the L18 FFDs incorporate the 2012 and 2015 epochs of the \sys\ data, they made up $<10$\% of the data, so will not significantly bias the comparison.
The flattening of the GJ~436 FFD at low energies represents the event detection limit.

\begin{deluxetable*}{cccccccccccc}
    \caption{FUV flares detected from GJ 436.}
    \tabletypesize{\scriptsize}
    \tablewidth{0pt}
    \tablehead{ \multicolumn{4}{c}{Summed Lines (Used to Identify Flares and in Figures \ref{fig:overview} \&\ \ref{fig:spi})} & \multicolumn{4}{c}{$FUV_{130}$ Band (per \citealt{loyd18b})} & \colhead{} & \colhead{} \\ \colhead{$\delta$} & \colhead{$E$} & \colhead{$F_{\mathrm{peak}}$} & \colhead{$\frac{F_\mathrm{peak}}{F_q}$\tablenotemark{a}} & \colhead{$\delta$} & \colhead{$E$} & \colhead{$F_{\mathrm{peak}}$} & \colhead{$\frac{F_\mathrm{peak}}{F_q}$\tablenotemark{a}} & \colhead{$t_\mathrm{peak}$} & \colhead{No.\tablenotemark{b}} \\ \colhead{s} & \colhead{$10^{28}$ erg} & \colhead{$10^{-14}$ $\mathrm{erg\ s^{-1}\ cm^{-2}}$} & \colhead{} & \colhead{s} & \colhead{$10^{28}$ erg} & \colhead{$10^{-14}$ $\mathrm{erg\ s^{-1}\ cm^{-2}}$} & \colhead{} & \colhead{MJD} & \colhead{} }
\startdata
$ 1514 \pm 130 $ & $ 3.76 \pm 0.26 $ & $ 4.85 \pm 0.67 $ & $ 23.2 \pm 6.0 $ & $ 745 \pm 120 $ & $ 3.76 \pm 0.49 $ & $ 6.70 \pm 0.82 $ & $ 16.1 \pm 5.5 $ & 58142.88 & 11\\
$ 1419 \pm 118 $ & $ 3.52 \pm 0.28 $ & $ 1.57 \pm 0.38 $ & $ 8.2 \pm 2.7 $ & $ 863 \pm 129 $ & $ 4.35 \pm 0.61 $ & $ 1.93 \pm 0.54 $ & $ 5.4 \pm 2.3 $ & 58143.15 & 12\\
$ 1337 \pm 112 $\tablenotemark{c} & $ 3.68 \pm 0.27 $ & $ 3.05 \pm 0.53 $ & $ 13.6 \pm 4.0 $ & $ 660 \pm 90 $ & $ 3.82 \pm 0.51 $ & $ 3.63 \pm 0.68 $ & $ 8.1 \pm 1.8 $ & 58110.89 & 7\\
$ 1018 \pm 117 $ & $ 2.29 \pm 0.24 $ & $ 4.77 \pm 0.66 $ & $ 25.1 \pm 5.2 $ & $ 594 \pm 111 $ & $ 2.81 \pm 0.52 $ & $ 5.99 \pm 0.81 $ & $ 15.4 \pm 2.7 $ & 58108.72 & 6\\
$ 1001 \pm 95 $ & $ 2.92 \pm 0.25 $ & $ 2.23 \pm 0.46 $ & $ 9.7 \pm 2.9 $ & $ 591 \pm 97 $ & $ 3.16 \pm 0.50 $ & $ 2.66 \pm 0.65 $ & $ 6.7 \pm 2.1 $ & 58177.33 & 13\\
$ 884 \pm 99 $ & $ 2.43 \pm 0.26 $ & $ 1.44 \pm 0.37 $ & $ 7.0 \pm 2.4 $ & $ 562 \pm 106 $ & $ 3.25 \pm 0.61 $ & $ 2.55 \pm 0.82 $ & $ 6.0 \pm 2.0 $ & 58111.42 & 8\\
$ 776 \pm 87 $ & $ 2.14 \pm 0.22 $ & $ 1.49 \pm 0.37 $ & $ 7.2 \pm 2.5 $ & $ 357 \pm 86 $ & $ 2.07 \pm 0.49 $ & $ 2.26 \pm 0.59 $ & $ 5.4 \pm 1.5 $ & 58111.70 & 9\\
$ 663 \pm 98 $ & $ 1.15 \pm 0.17 $ & $ 1.88 \pm 0.42 $ & $ 13.3 \pm 3.0 $ & $ 732 \pm 173 $ & $ 1.84 \pm 0.39 $ & $ 2.26 \pm 0.58 $ & $ 11.2 \pm 5.2 $ & 56101.31 & 1\\
$ 646 \pm 92 $ & $ 2.14 \pm 0.28 $ & $ 1.28 \pm 0.34 $ & $ 5.4 \pm 2.3 $ & $ 434 \pm 89 $ & $ 2.48 \pm 0.51 $ & $ 2.21 \pm 0.54 $ & $ 5.4 \pm 1.3 $ & 58137.65 & 10\\
$ 627 \pm 74 $ & $ 2.51 \pm 0.28 $ & $ 1.38 \pm 0.36 $ & $ 4.9 \pm 1.9 $ & $ 527 \pm 75 $ & $ 3.54 \pm 0.50 $ & $ 2.23 \pm 0.54 $ & $ 4.8 \pm 1.1 $ & 57199.12 & 5\\
$ 626 \pm 85 $ & $ 1.82 \pm 0.24 $ & $ 1.13 \pm 0.33 $ & $ 5.4 \pm 2.0 $ & $ 460 \pm 104 $ & $ 2.47 \pm 0.54 $ & $ 2.16 \pm 0.60 $ & $ 5.6 \pm 1.9 $ & 58177.46 & 14\\
$ 476 \pm 71 $ & $ 0.83 \pm 0.12 $ & $ 2.08 \pm 0.43 $ & $ 14.6 \pm 3.0 $ & $ 280 \pm 131 $ & $ 0.70 \pm 0.28 $ & $ 2.69 \pm 0.61 $ & $ 13.2 \pm 5.9 $ & 56101.37 & 2\\
$ 395 \pm 64 $ & $ 1.58 \pm 0.24 $ & $ 1.68 \pm 0.40 $ & $ 5.8 \pm 2.1 $ & $ 78 \pm 67 $ & $ 0.53 \pm 0.45 $ & $ 1.97 \pm 0.59 $ & $ 4.3 \pm 1.3 $ & 57199.10 & 4\\
$ 317 \pm 55 $ & $ 1.27 \pm 0.19 $ & $ 2.22 \pm 0.44 $ & $ 7.3 \pm 2.5 $ & $ 198 \pm 47 $ & $ 1.33 \pm 0.31 $ & $ 3.00 \pm 0.57 $ & $ 6.1 \pm 1.2 $ & 57198.99 & 3
\enddata

\tablenotetext{a}{Ratio of peak flux to quiescent flux.}
\tablenotetext{b}{Chronological order of the flare. Corresponds to the numerical labels in the figures.}
\tablenotetext{c}{Exposure began after the start of the flare as indicated by a lack of any data points below the quiescent level prior to the peak. Energy and equivalent duration are lower limits.}

    \label{tbl:flares}
\end{deluxetable*}

The GJ~436 FFD does not appear to follow the ``M~dwarf average'' power law from L18, exhibiting a cliff in the FFD toward large equivalent durations ($\delta$) and energies.
Within the cumulative exposure of 0.88~d, the equivalent duration and energy of the strongest expected event would be $\approx3000$~s and $\approx3\sn{29}$~erg (where the L18 FFD lines intersect the 0.88~d$^{-1}$ occurrence rate in Figure~\ref{fig:ffd}), but the strongest observed event is several times weaker.
The observed flare rate is not affected by exposure gaps since flares are effectively random in time \citep{wheatland00}. 
Observed energies can be underestimated when gaps cut off events (such as event 7), but the bias this introduces is below the statistical uncertainty resulting from a small sample size (see injection-recovery tests with gappy data in Appendix C of L18).

To test the possibility that the FFD cliff is a result of small sample size, we drew flares randomly from the L18 equivalent duration FFD for comparison.
In 900,000 trials, 10\% yielded no flares with equivalent durations beyond the cliff.
However, within this 10\%, most trials produced fewer total flares.
If we considered only trials that yielded at least as many flares with $\delta>500$~s as GJ~436 (180,000 trials yielding $\geq 8$ flares), only 0.02\% produced events no more extreme than those observed from GJ~436.
In other words, if GJ~436's FFD were consistent with other M~dwarfs, it is highly unlikely that it could yield as many flares as observed and yet produce none with equivalent durations $>860$~s.
This implies that GJ~436's FFD is steeper than typical.
A power-law fit to the 8 $\delta>500$~s events yields an exponent $<-2$, well below the $0.76\pm0.1$ typical of M~dwarfs (L18), although the fit is poor due to the low number of events. 
We hypothesize that this unusual FFD is the result of a magnetic SPI (\S\ref{sec:spi}).

To investigate whether GJ~436 can produce more energetic flares over longer timescales, we searched 43~d of optical observations by the Transiting Exoplanet Survey Satellite (TESS).
The presence of a flare in those data would have been evidence against the cliff in UV flare energies as a real feature.
However, we found no flares in the TESS data. 
Because flares produce very low contrasts at optical wavelengths \citep[e.g.,][]{macgregor21}, the possibility remains that flares with large UV energies occurred, but their optical signals were buried in the noise of the TESS time series.

\subsection{Rotation}
\label{sec:rotation}
\begin{figure*}
    \includegraphics{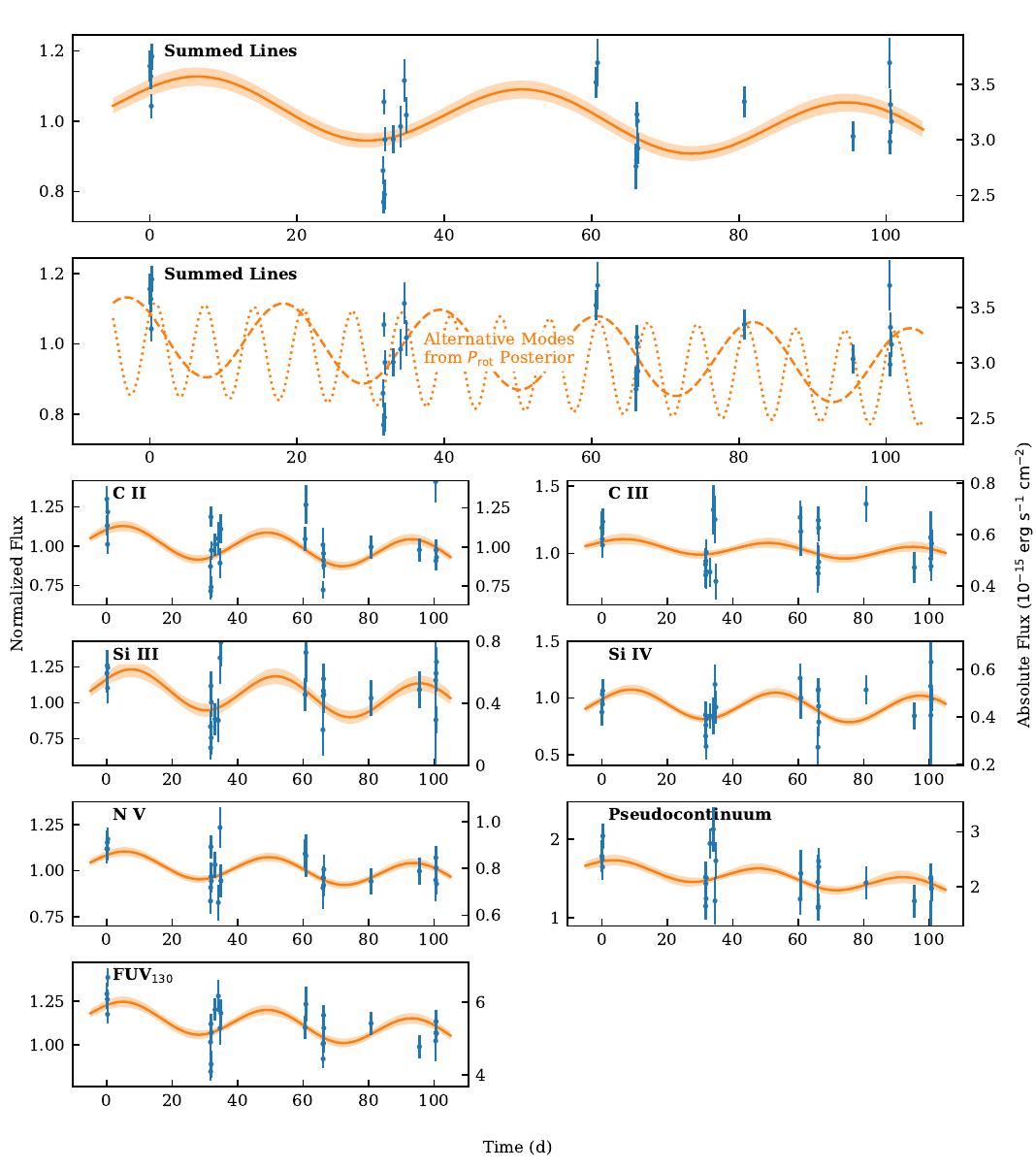}
    \caption{
    Fit to the rotational modulation of flux in each of the analyzed bands during the 2017/2018 epoch of observations.
    The data have been cleaned of flares.
    The fits were made to data binned to a $10$~min cadence, but to minimize clutter the plotted data are binned to full exposure length (typically around 45~min).
    Fits corresponding to alternative periods are shown as dotted and dashed lines in the panel second from top  (see text). 
    The slight downward trend in the UV fits is due to the simultaneous fit to the activity cycle. 
    }
    \label{fig:rotation}
\end{figure*}

Based on our fits to the data sampling 2.5 stellar rotation periods during the 2017/2018 epoch, we detected stellar rotational modulation at $>2\sigma$ in each band and 4.3$\sigma$ in the summed line flux, with typical amplitudes spanning 5-13\%.
Figure~\ref{fig:rotation} shows these fits and Table~\ref{tbl:fits} gives the fit parameters.
MCMC chains of the fits and corner plots are available upon request to the corresponding author.

\begin{deluxetable*}{lccccccc}
    \caption{Fits to variability in FUV emission lines.}
    \tablewidth{0pt}
    \tablehead{ \colhead{Band} & \colhead{Mean} & \colhead{Cycle} & \colhead{Cycle} & \colhead{Rotation} & \colhead{Rotation} & \colhead{SPI} & \colhead{Excess} \\ \colhead{} & \colhead{Flux} & \colhead{Amplitude} & \colhead{$t_\mathrm{0}$\tablenotemark{a}} & \colhead{Amplitude} & \colhead{$t_\mathrm{0}$\tablenotemark{a}} & \colhead{Amplitude} & \colhead{Noise\tablenotemark{b}} \\ \colhead{} & \colhead{$10^{-16}$ $\mathrm{erg\ s^{-1}\ cm^{-2}}$} & \colhead{\%} & \colhead{yr} & \colhead{\%} & \colhead{d} & \colhead{\%} & \colhead{\%} }
\startdata
Summed Lines & $31.92_{-0.74}^{+0.59}$ & $37.5_{-3.0}^{+3.3}$ & $3.905_{-0.075}^{+0.079}$ & $8.1_{-1.6}^{+2.1}$ & $ -2.1 \pm 1.5 $\tablenotemark{c} & $<$9.2 & $10.8_{-1.1}^{+1.5}$\\
C II & $10.01_{-0.29}^{+0.35}$ & $41.9_{-3.8}^{+6.2}$ & $ 3.86 \pm 0.10 $ & $9.5_{-2.2}^{+2.6}$ & $-3.7_{-1.1}^{+1.2}$ & $<$7.9 & $13.6_{-1.6}^{+2.4}$\\
C III & $5.25_{-0.27}^{+0.32}$ & $30.4_{-9.5}^{+9.2}$ & $4.04_{-0.31}^{+0.18}$ & $4.8_{-2.3}^{+3.1}$ & $-1.7_{-1.5}^{+1.4}$ & $<$13 & $<$12\\
Si III & $3.55_{-0.16}^{+0.22}$ & $50.8_{-9.5}^{+8.0}$ & $4.02_{-0.15}^{+0.13}$ & $13.0_{-3.7}^{+4.1}$ & $-1.2_{-1.3}^{+1.2}$ & $<$16 & $21.5_{-3.0}^{+4.1}$\\
Si IV & $4.74_{-0.21}^{+0.31}$ & $29.1_{-7.1}^{+8.7}$ & $3.66_{-0.42}^{+0.17}$ & $12.5_{-3.1}^{+3.0}$ & $0.1_{-1.2}^{+1.1}$ & $<$11 & $8.1_{-5.3}^{+3.6}$\\
N V & $7.89_{-0.27}^{+0.23}$ & $30.3_{-4.0}^{+6.8}$ & $3.95_{-0.15}^{+0.11}$ & $6.7_{-1.6}^{+1.8}$ & $-3.0_{-1.2}^{+1.1}$ & $<$4.8 & $6.9_{-3.6}^{+1.9}$\\
Pseudocontinuum & $14.4_{-5.1}^{+1.5}$ & $89_{-13}^{+174}$ & $4.538_{-0.065}^{+0.052}$ & $11.7_{-4.5}^{+7.8}$ & $-4.3_{-1.4}^{+1.8}$ & $<$10 & $14_{-8}^{+12}$\\
FUV$_{130}$ & $48.1_{-3.3}^{+1.8}$ & $47_{-8}^{+16}$ & $4.24_{-0.11}^{+0.06}$ & $8.3_{-2.0}^{+2.4}$ & $ -3.5 \pm 1.2 $ & $<$7.8 & $10.7_{-1.9}^{+2.1}$
\enddata

\tablenotetext{a}{Time at which phase of $\sin$ function reaches zero after 2018-01-01 00:00:00 UT (58119.0 MJD).}
\tablenotetext{b}{Additional white noise hyperparameter for the 2018 epoch, excluding flares.}
\tablenotetext{c}{Applied as a prior to remaining fits.}

    \label{tbl:fits}
\end{deluxetable*}

Taking into account the inclination of the star, the rotational variability can provide some insight on the net contrast of active regions. 
\cite{bourrier22} measured the star's spin axis to be inclined by ${35.7\mdeg}^{+5.9}_{-7.6}$ relative to the line of sight.\footnote{The flipped inclination of 144.2\deg\ is equally probable, but does not influence our interpretation.}
At this inclination, latitudes beyond $\pm$54.3\deg\ are always visible for one pole and invisible for the other. 
\cite{granzer00} found that active regions mostly occur within $\pm$54.3\deg\ for slowly rotating M~dwarfs.
Therefore, most of \sys's active regions likely contributed to the observed variability, albeit with an area foreshortened by the inclination.  

dS19 analyzed the same FUV observations for rotational modulations.
They found the rotational modulation of individual emission lines only to be significant if they included the 2015 observations under the assumption, justified by optical observations \citep{bourrier18}, that the signal phase did not shift between the 2015 and 2017/2018 epochs.
In our analysis, summing emission line fluxes resulted in a $>4\sigma$ measurement of the rotation signal amplitude without need of the 2015 data. 
Regardless, we consider rotational modulation of FUV emission as more likely present than absent, given signs of magnetic activity on \sys\ and rotational modulation of UV emission due to magnetic activity on the Sun \citep{toriumi20}.
The ds19 fits yielded rotational amplitudes of roughly 15\% in \Cii\ and \Siiii\ and 10\% in \Nv.
Our fits yielded lower levels of rotational variability (Table~\ref{tbl:fits}), likely because, in contrast with dS19, we did not include the higher fluxes of the 2015 epoch when fitting for rotation due to concern that phase and amplitude could have shifted between epochs.

We tested whether the FUV data could recover the rotation period of the star if we included rotation period as a free parameter in the variability fit.
In this case, the MCMC sampler identified multiple peaks in the posterior for rotation period, with the two strongest at $21.3_{-1.1}^{+0.2}$~d and $6.69\pm0.04$~d (Figure~\ref{fig:rotation}, second panel) and only a weak peak in the vicinity of 44~d.
Amplitudes of the short-period signals were greater than the fixed-period fit, but within 2$\sigma$. 
The multiple modes could be harmonics of the longer optical period, perhaps indicating that multiple longitudes of enhanced activity are contributing to the FUV variability.

In comparison to FUV emission, variability in broadband optical emission is $\sim$0.1\%.
Our sinusoidal fit to the optical data (Figure~\ref{fig:lothringer}) yielded a rotation period of $44.09\pm1.16$~d with zero phase occurring at JD $2455484.47\pm0.73$ (decimal year $2010.7862\pm0.0020$) and an amplitude of $0.127\pm0.013$\%, consistent with the results of \cite{bourrier18} and \cite{lothringer18}.
In contrast, rotation measurements made using \Caii~H~\&~K and \Ha\ suggest periods nearer to 40~d \citep{suarez15,kumar23}.
A rotation period within the 40-45~d range is typical of a field-age M~dwarf with a mass $\approx0.45$~\Msun\ \citep{bourrier18,newton18}.
If the rotational variability is due to dark spots, the star should appear redder during optical lows, but \cite{lothringer18} could not detect this, suggesting that the variability could be dominated by faculae.

The phase difference between the UV and optical rotation curves encodes information about the surface structure of the stellar activity.
If the rotation signals result from optically bright faculae cospatial with FUV plage, then the two should be in phase.
Alternatively if optically dark spots cospatial with FUV plage are responsible, then the two curves should be 180\deg\ out of phase.

In this case, however, a clear comparison of the UV and optical phases is precluded by the possibility of phase shifts in the optical rotation signal.
The optical variability fit assumes a static phase over the 14 years of data, a span that covers two activity cycles.
Although the data, folded onto the best-fit period, show convincing modulation (Figure~\ref{fig:lothringer}), hidden phase shifts over the 14 year span could be present, particularly if the signal is dominated by only one or a few short periods of high variability.  

To investigate possible shifts in the optical phase, we isolated 3-year sections of the optical data sampling extrema of the activity cycle (2006-2008, 2010-2012, and 2013-2015).
Fits to data near optical minima recovered the same 44~d period, with a larger amplitude near the 2014 minimum (0.27 vs. 0.13\%) and large uncertainties in phase allowing for shifts as high as 123\deg\ (1$\sigma$) from 2007 to 2014 that could affect an FUV-optical comparison. 
The fit to data near the 2011 maximum did not clearly recover the 44~d period, yielding multiple modes in period and phase, all with amplitudes near 0.9\%.
The unclear period and lower variability amplitude near the 2011 maximum could result from a lower activity level (fewer or weaker spots/faculae), a more homogeneous distribution of spots/faculae across the star, cancellation of spots by faculae, or any combination of the above.
Since the 2017/2018 FUV observations occurred near the subsequent optical maximum, they could also be affected by reduced or more spatially homogeneous activity.

We also attempted fitting TESS data from the two available visits, sectors 22 and 49 observed in 2020 March and 2022 March, each lasting only about half of the stellar rotation period.
Separate fits to each did not reveal a clear rotation signal or phase constraint.
Higher precision optical data are necessary to probe signal shifts over time.

\cite{toriumi20} provides solar context for how rotational variations, resulting from spots and faculae, manifest at optical versus UV wavelengths.
Isolating periods where only a single active region was visible, they found variability in extreme UV (EUV) emission lines can be either in or out-of-phase with optical variability.
The determining factor is whether the  brightened active region or its dimmed surroundings dominate.
The dimming is caused by heating of surrounding plasma to higher temperatures, reducing emission from lower temperature lines.
Variations in EUV emission also start earlier and end later than optical variations because EUV emission originates above the photosphere, making it visible when photospheric active regions are hidden just beyond the limb. 
Finally, EUV variations exhibit top-hat-like shapes due to low optical depths.
GJ~436's FUV variability could exhibit similar effects, though, unlike the transition-region FUV lines we analyzed, the most analogous EUV lines analyzed by \cite{toriumi20} include substantial flux from hotter, less dense coronal plasma.

\subsection{Activity Cycles}
The activity cycle of GJ~436 was measured by \cite{lothringer18} at optical wavelengths to have a period of roughly 7.4~yr with an amplitude of 5~mmag (0.5\%).
Our analysis of the same data yielded a period of $7.75\pm0.10$~yr and an amplitude of $0.376\pm0.015$\% (4~mmag), with zero phase occurring at JD $2454826\pm18$ (decimal year $2008.982\pm0.049$).
A long term increase of $0.050\pm0.003\%$ over the 14 year span of observations is also present. 
The 0.35~yr difference between our cycle period and that of \cite{lothringer18} is likely a systematic uncertainty due to subjective analysis choices (using flux vs. magnitude, functional form of the long term trend, functional form of the cycle variability, inclusion and form of a jitter term, etc.). 
An additional search for the star's activity cycle signature in \Caii~H~\&~K, \ion{Na}{1}, and \Ha\ activity indicators by \cite{kumar23} yielded signals spanning 5.1-6.8~yr.
All of the aforementioned values fall near the predicted value of 7.6~yr based on the relationship between activity cycle and rotation period established by \cite{suarez16}.
The simple sinusoidal shape of the activity cycle echoes that of field-age Sun-like stars with weak, faculae-dominated activity.

The three epochs of HST data span nearly a full activity cycle and show significant variability beyond what can be explained by stellar rotation, visit-to-visit systematic flux errors ($<2$\%, \citealt{james22}), or excess noise (2-4\% when binned over a 2-5~orbit visit).
Fitting a sinusoid to these data, fixed to the period of the optical cycle, yields amplitudes of 30-50\%. 
The amplitude of cycle variability in the pseudocontinuum stands out at $89^{+174}_{-13}$\%.
We interpret this difference with caution because, on a pixel level, flux in the pseudocontinuum was below COS's noise floor and vulnerable to errors in the background subtraction and time-varying flat-field correction. 

\begin{figure}
    \includegraphics{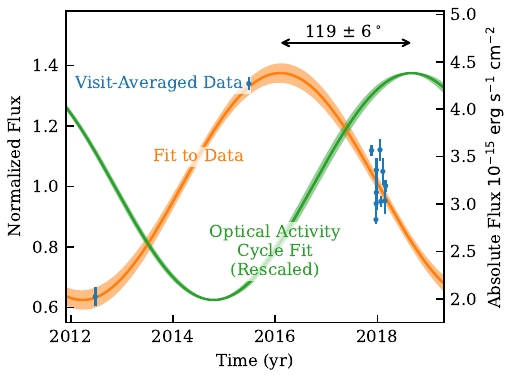}
    \caption{
    Summed line flux (blue points), binned by HST visit, across all three epochs of HST data, fit with a sinusoidal model for activity cycle variations (orange curve) fixed to the optical cycle period, and shown alongside the (rescaled) fit to optical cycles (green curve) to compare phases. 
    }
    \label{fig:cycles}
\end{figure}

Similar to rotational variability, if optically-dark, FUV-bright active regions explain this variability, then the optical and FUV activity cycles should be 180\deg\ out of phase \citep{reinhold19}.
For \sys, optical variability lags the FUV variability by $119\pm6$\deg (Figure~\ref{fig:cycles}).
dS19 found that variations in the \Caii\ S-index were out of phase with optical variations, though the constraint was not quantified.

The intermediate FUV-optical phase offset could indicate that \sys\ is undergoing a transition from spot-dominated to faculae-dominated activity.
Sun like stars exhibit similar intermediate phase offsets between their activity cycles in \Caii~H~\&~K and optical emission as they transition from spot to faculae dominance \citep{reinhold19} around a Rossby number of one, similar to \sys's Rossby number of 0.74 \citep{loyd21}.
A transitional state for \sys\ could explain its mixed indications of spot and faculae-driven variability.

Systematic unknowns could make a 180\deg\ FUV-optical phase difference possible.
A global trend of about 7\%~\peryr\ to the FUV emission would suffice, as would allowing the FUV cycle to vary in amplitude such that it is about half the value in 2018 as in 2012.
Another option is to interpret the 2012 and 2015 observations as sampling a rotational minima, meaning average fluxes were higher at those epochs, which would shift the cycle curve to the earlier times. 
The star's increase in optical variability near 2015 (Section~\ref{sec:rotation}) could indicate larger FUV variations at that time.
On the Sun, the amplitude of rotational variability waxes and wanes over activity cycles at optical and EUV emission wavelengths (\citealt{frohlich06,woods22}; Llama, private comm.).

\subsection{Excess White Noise}
We included an excess white noise parameter (``jitter'' term) in our fits so that the MCMC sampler could explore an alternative to forcing rotational, cyclic, or SPI signals to produce the observed point scatter.
The sampler identified excess white noise in addition to other variability at levels of 10-20\% in all bands except \Ciii.
Amplitudes of excess noise and other variability were uncorrelated, i.e., increasing the noise did not require decreasing other variability to obtain a similar goodness of fit. 

Excess noise could come from a mixture of instrumental and astrophysical sources.
A known instrumental source is the movement of fixed pattern noise when the position of the spectrum on the detector is shifted (as much as 36~\AA, or 10\% of the wavelength range) between observations.
However, instrumental noise of this kind is unlikely to account for the entirety of the measured excess noise given the 2\% visit-to-visit accuracy of COS \citep{james22}.
Astrophysical sources could include undetected flares, ``transition region bombs,'' shocks initiated by convective motions, and others \citep{loyd14}.

Excess noise, instrumental or astrophysical, limits the detectability of planet b's transit in FUV emission lines aside from \lya\ (\citealt{loyd17, lavie17}; dS19).
Integrated to a 1~h cadence (and assuming a flat power spectral density), the noise level is $5.8\pm0.8$\% in emission from the least-ionized ion we analyzed, \Cii.
\cite{loyd14} estimated excess noise with an analytical method and a 60~s cadence, using only the 2012 data, and found values equating to $2.5\pm1.4$\% in \Cii\ at a 1~h cadence.

\subsection{Erratum: Corrected magnetic field limit for GJ 436 b}
This section reflects an erratum relating to the subsequent section submitted on 2024 February 23 and has been copied here for the convenience of readers of the arXiv version. 

In the original manuscript, it was reported that the upper limit on variations in far ultraviolet line emission from GJ~436 phased with the orbit of planet b implied an upper limit of $\sim$100~G. This value is incorrect due to an arithmetic error discovered in the analysis scripts following publication. The revised limit is $\sim$10~G. In contrast with the previous value, the revised limit is comparable to the magnetic field strength of Jupiter and lower than the magnetic fields reported for 4 hot Jupiter planets using the same methodology with \Caiihk\ observations \citep{cauley19}.

Later work by \cite{vidotto23} has shown that an assumption critical to this estimate is invalid. The estimate is based upon the model of \cite{lanza13}, which assumes the planet orbits in a potential magnetic field. The \cite{vidotto23} model demonstrated that \planet\ orbits beyond the range where the stellar field is approximately potential. Although invalid for \planet\ specifically, our field constraint remains generally useful for estimating the sensitivity of FUV observations of magnetic SPI to exoplanetary magnetic field strengths. Note that the hypothesis that the unusual flare behavior of \sys\ might be a result of a magnetic star-planet interaction remains viable, as the \cite{vidotto23} model confirmed that \planet\ primarily orbits within the \alf\ surface of the star.

\subsection{Magnetic Star-Planet Interaction}
\label{sec:spi}
To search for a star-planet interaction, our HST program (HST-GO-15174) was designed to add 8 points sampling across the orbital phase of GJ~436 b.
The data augmented the transit investigations of program HST-GO-14767.
This sampling was motivated by a peak in the \Nv\ emission of GJ~436 in phase with the planet transit during the 2015 epoch.

\begin{figure*}
    \includegraphics{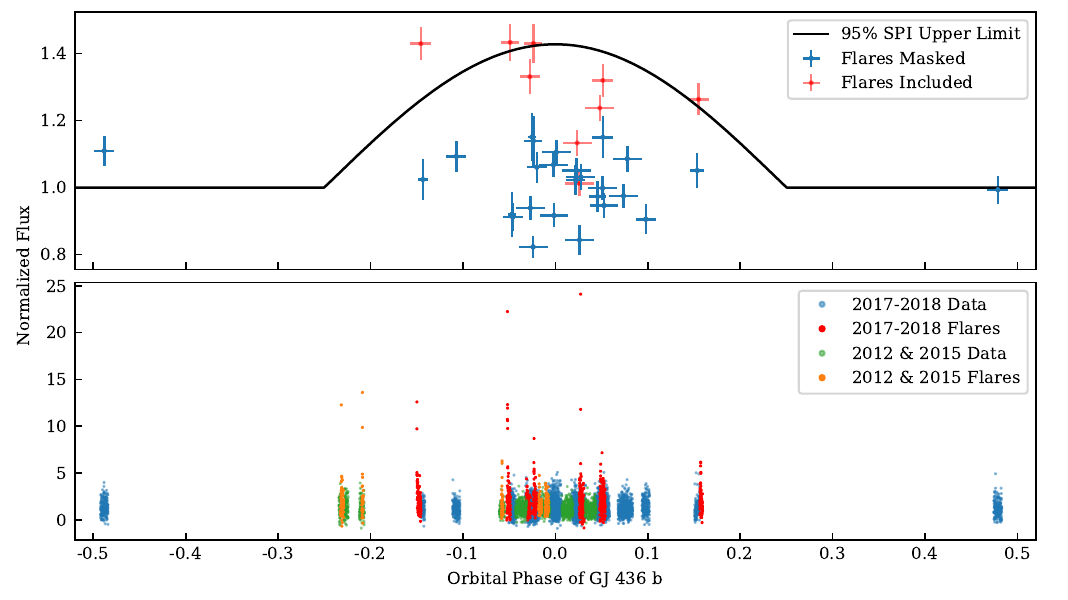}
    \caption{
    Fluxes plotted against the planetary orbital phase.
    The planet's optical transit spans a narrow range of $\pm0.008$ in phase.
    Top: Data from the 2017/2018 epoch binned by exposure with the best-fit rotational modulation subtracted.
    Blue points exclude and red points include flares.
    The black solid line shows the $2\sigma$ upper limit amplitude for a truncated sine SPI model fit to the data with flares included.
    Bottom: Data at a 10~s cadence to show individual flares.
    Blue (quiescent) and red (flare) data are from the 2017/2018 epoch whereas green (quiescent) and orange (flare) data are from the earlier 2012 and 2015 epochs.
    We speculate that the lack of flares when the planet is near eclipse (phase 0.5)  could suggest a magnetic SPI is triggering the star's frequent small flares, but with energies too weak to register as a detection using the truncated sine model. 
    }
    \label{fig:spi}
\end{figure*}

We found no detectable evidence of an SPI in the form of variation in the emission from GJ~436 in \Nv\ or any other line phased with the planetary orbit.
Our simultaneous fits limit SPI variability in summed line flux to $<9.4\%$ (2$\sigma$).
The MCMC sampler strongly preferred sinusoidal solutions to a top hat.
In Figure~\ref{fig:spi}, we plot a model set at the $2\sigma$ upper limit on the SPI amplitude against the data, after subtracting the best-fit rotational modulation signal from those data. 

The upper limit on the SPI signal enabled us to place a corresponding limit on the strength of the planetary magnetic field, under certain assumptions:

An initial assumption is that magnetic disturbances from the planet propagate back to the star.
For this to occur, the \alf\ Mach number at the planet must be $<1$ \citep{lanza15}.
\cite{saur13} estimated the \alf\ Mach number of \sys's wind near \planet\ to be 1.01, very near the limit.
This value relied on an empirically-scaled estimate of the stellar wind speed at the planet of 235~km~\pers.
In contrast, \cite{bourrier16} estimate a lower value of 85~km~\pers\ based on matching a model of the planetary outflow's dynamical interaction with the stellar wind to \lya\ transit data.
This would imply an \alf\ Mach number $\approx0.3$, making a magnetic SPI capable of propagating to the star.
However, this still relies on an estimate of the stellar magnetic field based on a scaling relationship for Sun-like stars \citep{saur13}.
A direct measurement of \sys's magnetic field is necessary to better gauge whether \planet\ orbits within a region where the \alf\ Mach number is $<1$.

A second assumption is that the interaction takes the form of the ``flux tube dragging'' scenario set forth by \citep{lanza13}.
In this scenario, a persistent magnetic flux tube links star and planet.
The orbital motion of the planet drags this tube through the ambient stellar magnetic field, triggering the stellar field to relax to lower energy states, releasing energy in the process.
This is the only SPI configuration that predicts energy release at a level consistent with past observations of stellar activity-related emission that is modulated at the orbital period of a close-in planet \citep{cauley19}.
Note that for other SPI models, the relative orientation of the planetary and stellar fields can play a critical role \citep[e.g.,][]{strugarek16}.

We followed the methodology of \cite{cauley19} to estimate an upper limit on \planet's magnetic field.
The method requires estimates for a number of quantities:
For the stellar field at the planet, we took the value of 667~nT as estimated by \cite{saur13}.
We set the relative velocity between the planet and the star's magnetic field to be the same as the mean orbital velocity, estimated from the period and semi-major axis found by \cite{lanotte14}.
This simplification is justified because the planet's orbital velocity is over an order of magnitude larger than the rotational velocity of the stellar magnetic field at the orbital distance of the planet.
We used the planetary radius value of $4.10\pm0.16$~\Rearth\ from \cite{lanotte14}.
Finally, to estimate the fraction of SPI power radiated in the summed FUV line flux, we used the flare energy budgets from panchromatic observations of stellar flares on the active M~dwarf AD~Leo reported by \cite{hawley03}.
These indicate that roughly 3\% of the total energy emitted in a flare is emitted in the five lines that we summed.
From the above values, we estimated an upper limit on the planetary magnetic field of $\lesssim$100~G.
This limit is well above Jupiter's magnetic field strength and within the range of values \cite{cauley19} estimated for several hot Jupiters.

Besides inducing a hot spot, a close-in planet could affect the activity level of its host in several ways.
It could suppress activity by tidally disrupting the stellar convective dynamo \citep{pillitteri14,fossati18}.
It could obscure activity by enshrouding the star in absorbing gas from a radiation-powered planetary outflow \citep{fossati13,staab17}.
Alternatively (or even concurrently), it could enhance activity by tidally inhibiting the spin-down of the host star \citep{poppenhaeger14,tejada21}.
\sys's activity, in the form of FUV emission lines, falls a factor of a few below the predicted level for an M star with a 44~d rotation period \citep{loyd21,pineda21}, but is not an outlier.
For comparison, the early M star GJ~832 has a slightly shorter rotation period ($37.5\pm1.5$~d; \citealt{gorrini22}) and a similar level of FUV emission, yet RV surveys have not revealed a close in, massive planet \citep{bailey09,gorrini22}. 
We conclude that \sys's level of FUV emission is not strongly influenced by the presence of its warm Neptune, but that a magnetic SPI could nonetheless be present. 

Although we did not detect an SPI in the form of orbit-modulated emission, we speculate that \sys's odd FFD could indicate a magnetic SPI. 
The enhancement of relatively low energy (equivalent duration $\approx$500~s flares) and statistically significant lack of more energetic events could indicate that a magnetic SPI is triggering frequent releases of magnetic energy at low levels, thereby preventing the build up of stored energy necessary to give rise to the larger flares exhibited by other M~dwarfs.

The comparison sample from L18, which produced stronger flares, are less prone to a magnetic SPI.
Within the sample, 8/10 (6/6 in the inactive sample) are known to host planets (including GJ~436, see Section~\ref{sec:flares}).
None are as massive and as close-in as \planet, though several come to within about a factor of two of mass or distance.
The closest comparison is GJ~581~b, a planet with $M \sin i = 15.8\pm0.3$~\Mearth\ orbiting 0.040~AU from its host \citep{robertson14} versus \planet's  $25.4^{+2.1}_{-2.0}$~\Mearth\ mass and 0.031~AU distance from its host \citep{lanotte14}. 
For the two stars without known planets, RV limits do not exclude planets like GJ~436~b \citep{bailey12,kossakowski22}.

Consistent with an SPI hypothesis is the lack of flares during planetary eclipses, when a sub-planetary hot spot on the star would be most likely to be invisible.
The eclipse data are too brief for this difference to be statistically significant.

Theoretically, the increase in time-averaged emission caused by SPI-triggered flares should also produce a detectable orbit-modulated signal.
For this reason, we included the flux added by flares when fitting for a possible SPI signal. 
With flares included, the MCMC sampler does favor a slight increase in emission within the $-0.25$ to 0.25 phase range, but it is marginal, with a likelihood ratio near unity.
Within this phase range, the average flux (after removing the best-fit rotation and cycle variability) is 1.7$\sigma$ higher than the two visits near the planetary occultation.
With more data, changes in the flare rate with planet phase, variations in time-averaged emission with planet phase, and FFD statistics could all lead to confirming or refuting an SPI in this system. 
The present data hint that flare statistics might have the greatest quantitative power in identifying a magnetic SPI. 

\subsection{Variability Synopsis}
\begin{figure*}
    \includegraphics{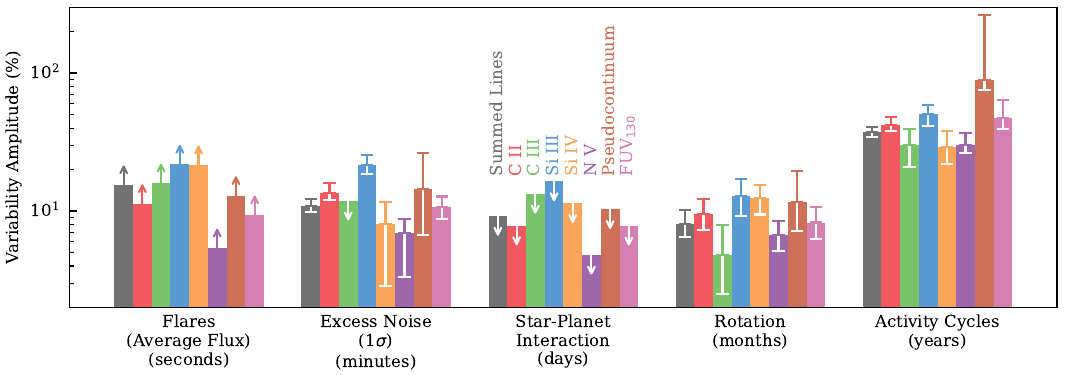}
    \caption{
    Summary of the contributions from various sources to the FUV variability of GJ 436.
    Arrows indicate upper or lower limits (see text).
    In parenthesis below each label are the timescale at which we measured each form of variability.}
    \label{fig:summary}
\end{figure*}

To place each form of variability in context, we compare their amplitudes side-by-side in Figure~\ref{fig:summary}.
Each type of variability has its own time structure, so each type has its own definition of ``amplitude.'' 

Flares do not have a well defined amplitude since they occur stochastically with large variations in peak flux (Figure~\ref{fig:ffd}).
As a metric for comparison, we adopted the time-averaged contribution of flares to the star's emission.
This number represents a lower limit on the true contribution of flares to the star's time-averaged emission because it omits both more frequent, weaker flares below the detection limit and larger, rarer flares not caught within the limited duration of the observations. 

Excess noise, since we assume it to be white noise, does not have time structure.
However, better data could reveal structure on timescales of seconds to minutes that the present data do not resolve. 
As a metric for comparison, we adopted the 1$\sigma$ level of the noise.

Our models for variability due to rotation, activity cycles, and SPI all have well-defined amplitudes. 
Since we did not detect SPI variations, Figure~\ref{fig:summary} shows only the $2\sigma$ upper limits on the amplitudes of the fits.

Activity cycles appear to be responsible for the greatest share of the variability observed across the observations.
If the Sun is any guide, activity cycles will not only affect the average level of FUV emission, but also the frequency of flares and the amplitude of rotational variability.
In other words, the amplitude of other variability sources will wax and wane over the course of the stellar activity cycle. 

\begin{figure*}
    \includegraphics{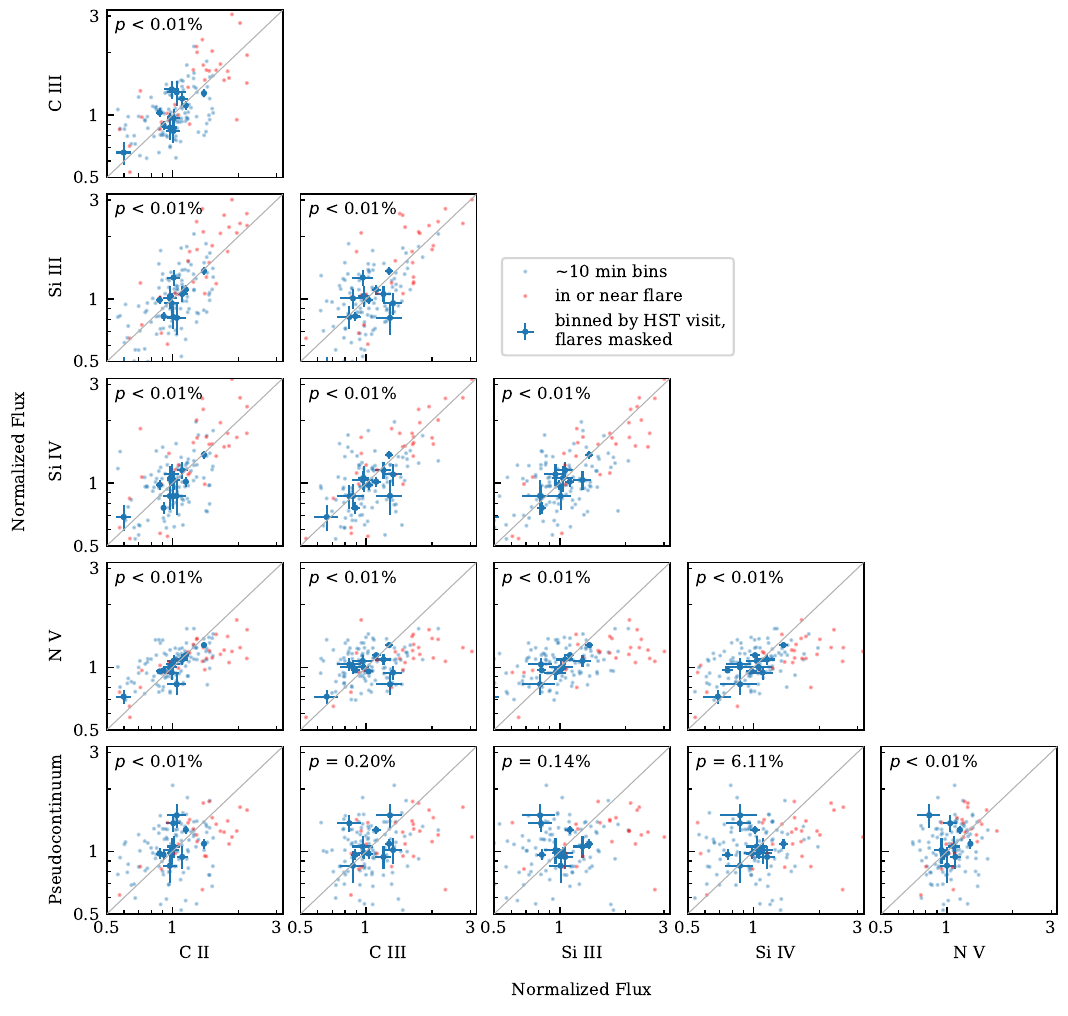}
    \caption{
    Correlations between the variability in each of the analyzed bands.
    The \fuv\ and Summed Line bands are omitted since they are a compendium of all the other bands (thus not independent).
    Gray lines indicate a 1:1 correlation.
    The $p$-value from a Spearman rank-order correlation test is given in each panel.
    Most band pairs are correlated to very high confidence, with the exception of some correlations with the pseudocontinuum.
    The largest excursions from the means are driven by flares, indicated by red coloring of the points.
    Notably, neither \Nv\ nor the pseudocontinuum seem to follow a 1:1 relationship with other bands, although the statistical significance of the relationship remains very high despite its shallower slope.}
    \label{fig:correlations}
\end{figure*}

Variability in most of the FUV bands we considered is well correlated.
Figure~\ref{fig:correlations} plots the correlation between the fluxes of each pair of bands.
For most line pairs, the points are consistent with a 1:1 correlation.
The notable exceptions are \Nv\ and the pseudocontinuum.
Variations in these bands, particularly at the high end (mostly due to flares), appear subdued relative to other lines, in line with observations of M dwarf flares (\citealt{france16}, L18, \citealt{france20}). 
Despite having a shallower slope for \Nv\ and the pseudocontinuum, correlations are still present at high significance in all cases except the correlation between the pseudocontinuum and \Siiv. 

We expect the extreme UV (EUV) emission of GJ~436 to track the changes in FUV line fluxes.
\cite{bourrier21} conducted differential emission measure modeling of the similar early M1.5 dwarf GJ~3470 across several observation epochs.
From 2018 to 2019, GJ~3470's FUV lines fluxes increased by about 30\%, while the flux in a 100-920~\AA\ band predicted by the DEMs increased by about half that, or 15\%.
In contrast, the Sun's emission variability across activity cycles increases toward shorter wavelengths \citep{woods22}, suggesting the opposite trend: that variations in the EUV should exceed those in the FUV for \sys.
\cite{france18} also noted an effectively linear correlation between \Siiv\ and \Nv\ emission measured by HST and 90-360~\AA\ flux measured (at a different epoch) by the Extreme UltraViolet Explorer (EUVE) for a sample of 11 F-M~dwarfs.
Ultimately, simultaneous FUV-EUV observations of \sys\ or similar stars are needed to quantify the relationship between FUV and EUV variability, but, at least qualitatively, EUV variations will track FUV variations. 
We leave DEM or stellar modeling reconstructions of GJ~436's EUV emission to future work.  

\section{Summary}
We conducted a comprehensive variability analysis of the far-ultraviolet (FUV) emission from the M2.5V exoplanet host star GJ 436.
Our analysis addressed flares, rotation, activity cycles, catch-all ``excess noise,'' and a possible star-planet interaction (SPI), with comparison to the optical variability measured by \cite{lothringer18}.
We focused on the summed emission from the \Cii, \Ciii, \Siiii, \Siiv, and \Nv\ lines in the 1150-1450~\AA\ range, the strongest lines in this range aside from \lya\ and \Oi, which were contaminated by geocoronal airglow.

Figure~\ref{fig:summary} provides a rapid comparison of the contribution of each variability source to the star's overall variability. 
The star's 2012-2018 activity cycle was responsible for the greatest degree of observed variability, $38\pm3$\% in amplitude.
Observed flares increased the time-averaged emission by 15\%.
Flares not observed, either because they were too weak to register above the noise or too rare to be captured in the limited observing time, would add to this value.
Stellar rotation produced variations of $8.1^{+2.1}_{-1.6}$\% and a mixture of instrumental and astrophysical sources treated with a catch-all white noise term account for an additional $10.8^{+1.5}_{-1.1}$\% variability in excess of photon noise. 
We did not detect variability phased with the planet's orbit, with a $2\sigma$ upper limit of 9.2\%.

Uncertainty in the optical rotational phase at the time of the FUV observations precludes strong conclusions about the relative distribution of spots, faculae, and plage.
Some other factors suggest the star could be undergoing a transition from spot- to faculae-dominated activity, including the lack of reddening during optical minima found by \cite{lothringer18}, the sinusoidal form of the star's optical activity cycle, a possible $119\pm6$\deg\ phase offset between the optical and FUV cycles, and the star's Rossby number near unity. 

If the planet is magnetically interacting with its star, its magnetic field must have a strength $\lesssim100$~G.
Our upper limit on the SPI signal assumes variations appear as a truncated sine function, such as would be produced by a planet-induced hot spot traversing the stellar surface.
However, the star produced an unusual population of flares that we conjecture could be the result of a magnetic SPI.
Although the flares were numerous, totaling 12 (plus 2 dubious) events over the course of about a day of cumulative exposure, their energies were anomalously low in comparison to other M~dwarfs.
We speculate that this could result from a magnetic SPI triggering a multitude of flares with low equivalent durations (a metric of flare energy normalized by the star's quiescent flux).
This would increase the flare rate at low equivalent durations while disrupting the energy buildup necessary for stronger events.
The lack of flares during planetary eclipse support this view, but these visits were too brief to have statistical power. 
\\
\\
\\
The authors thank the anonymous referee for their thoughtful review and suggestions, as well as Antonio Lanza and Sebastian Pineda for helpful discussions that improved this work. 
ROPL conducted this research under program HST-GO-15174.
Support for program HST-GO-15174 was provided by NASA through a grant from the Space Telescope Science Institute, which is operated by the Associations of Universities for Research in Astronomy, Incorporated, under NASA contract NAS 5-26555.
This research is based on observations made with the NASA/ESA Hubble Space Telescope obtained from the Space Telescope Science Institute, which is operated by the Association of Universities for Research in Astronomy, Inc., under NASA contract NAS 5–26555. 
These observations are associated with program(s) 12646, 13650, 14767, and 15174.
\\
\\

\software{emcee \citep{foreman13}, calCOS (https://github.com/spacetelescope/calcos), Astropy \citep{astropy13, astropy18, astropy22}, NumPy \citep{harris20}, SciPy \citep{virtanen20}}

\end{document}